\documentclass{article}
\usepackage[utf8]{inputenc}
\usepackage{amsmath}
\usepackage{hyperref}
\usepackage{graphicx}
\usepackage{todonotes}
\usepackage{fancyvrb}
\PassOptionsToPackage{hyphens}{url}\usepackage{hyperref}
\usepackage{bbm}
\usepackage[toc,page]{appendix}

\usepackage[style=numeric, citestyle=numeric-comp, sorting=none]{biblatex}
\usepackage{amsfonts}

\usepackage[left=.75in,top=.75in,right=.75in,bottom=.75in]{geometry}
\usepackage{setspace}

\title{Evaluating shifts in mobility and COVID-19 case rates in U.S. counties: A demonstration of modified treatment policies for causal inference with continuous exposures}
\date{2021}

\begin{document}
\maketitle

\begin{abstract}

\noindent

\textbf{Background:} 
Previous research has shown mixed evidence on the associations between mobility data and COVID-19 case rates. Evaluating the potential impact of decreased mobility on COVID-19 transmission is important, but complicated by differences between places on factors influencing both behavior and health outcomes.\\

\textbf{Methods:} We aimed to evaluate the county-level impact of shifting the distribution of mobility on the growth in COVID-19 case rates from June 1 - November 14, 2020. To do so, we utilized a \emph{modified treatment policy} (MTP) approach, which considers the impact of shifting an exposure away from its observed value. The MTP approach facilitates studying the effects of continuous exposures while minimizing parametric modeling assumptions. Ten mobility indices were selected to capture several aspects of behavior expected to influence and be influenced by COVID-19 case rates: Facebook’s proportion of residents staying at home and number of locations visited; PlaceIQ’s density of people at locations visited; Descartes Lab’s median and maximum distances traveled; and Google’s indices of retail/recreation business visits, workplace attendance, public transit use, and time spent at home. The outcome was defined as the number of new cases per 100,000 residents two weeks ahead of each mobility measure. Primary analyses used targeted minimum loss-based estimation (TMLE) with a Super Learner ensemble of machine learning algorithms, considering 20+ potential confounders capturing counties’ recent case rates as well as social, economic, health, and demographic variables. For comparison, we also implemented unadjusted analyses.\\

\textbf{Results:} Analyses included 1,182 of the most populous U.S. counties, covering 90\% of the population. For most weeks considered, unadjusted analyses suggested strong associations between mobility indices and subsequent growth in case rates. However, after confounder adjustment, none of the indices showed consistent associations after hypothetical shifts to reduce mobility. Sensitivity analyses varying the confounder set and time lag between exposure and outcomes yielded similar results.\\

\textbf{Conclusion:} Application of MTP facilitated examination of associations between reductions in mobility and subsequent COVID-19 cases in U.S. counties while minimizing parametric assumptions. After adjusting for measured confounders, common mobility metrics were generally not associated with new COVID-19 cases in Summer and Fall of 2020. While identifiability concerns limit our ability to make causal claims in this analysis, MTPs are a powerful and underutilized tool for studying the effects of continuous exposures.\\
\end{abstract}


\section{Background}

During a pandemic caused by a virus novel to the human immune system, such as today's SARS-nCoV-2 (COVID-19), policymakers have only non-pharmaceutical interventions (NPIs) to rely upon while treatments and vaccines are developed, tested, and distributed. Many NPIs are designed to reduce the number of close contacts a person has in order to curtail community spread. However, NPIs are only effective if they actually create the desired changes in behavior and if the changes in behavior have a causal impact on disease transmission.

To aid in COVID-19 response, private companies and research groups \cite{huang_twitter_2020, pepe_covid-19_2020, couture_measuring_2020, gao_mapping_2020} developed and maintain data sets aiming to quantify mobility and social distancing in the U.S. and worldwide \cite{google_covid19_nodate, facebook_movement_nodate}. These mobility indices are generated from anonymized mobile device tracking data aggregated at some geographic level, such as ZIP code or county. Despite privacy concerns, these indices may be a useful tool for researchers and public health officials \cite{buckee_aggregated_2020}. However, it is not clear which of these indices, if any, are meaningful proxies for behaviors that are related to COVID-19 spread, and whether any associations identified persist over time \cite{grantz_use_2020}.

Our aim was to evaluate the potential causal impact of mobility on new COVID-19 case rates, a goal complicated by the presence of confounders, varying epidemic arrival time across different areas, and changing governmental policies. A brief review of the literature on mobility and COVID-19 transmission is provided in Appendix \ref{appendix:lit}; our work contributes to that corpus in four critical ways. First, we examined the relationships between mobility and COVID-19 case rates over a long (four month) timeframe. Second, we employed a \emph{modified treatment policy} (MTP) approach, described next, to define relevant causal effects and avoid arbitrary categorizations of the continuous exposure. Third, we used targeted minimum loss-based estimation (TMLE) with Super Learner for estimation and inference, allowing us to adjust for a rich set of county-level confounders while relaxing parametric assumptions used in much of the other modeling research. Finally, we examined a wide set of mobility indices to determine which, if any, might be associated with COVID-19 case rates, and hence be possible proxies for the underlying behaviors that drive COVID-19 spread. 

All of our exposure data was continuous. Given the extensive literature on causal inference for binary exposures, one approach to examining mobility effects would be to discretize the data into ``high'' / ``low'' levels and ask what would be the expected difference in counterfactual COVID-19 case rates if all counties had that level of mobility. However, selecting a binary cut point for this categorization is arbitrary and risks losing information. Additionally, counties with a high proportion of essential workers might not be able to reduce their overall mobility to the specified ``low" level \cite{bembom_practical_2007}, violating the positivity assumption that all counties must have nonzero probability of receiving the specified exposure level within all values of measured adjustment variables \cite{petersen_diagnosing_2012}. As an alternative to categorizing the exposure, we could target the parameters of a working marginal structural model \cite{robins_marginal_1998, neugebauer_nonparametric_2007} or directly attempt to estimate a causal `dose-response' curve to capture how a continuous measure of mobility impacts the expected counterfactual case rates \cite{westling_causal_2019, kennedy_non-parametric_2017}. However, as before, there may be challenges with data support from structural and practical violations of the positivity assumption.

Instead, we took non-parametric approach to specify a causal effect that minimized parametric assumptions and naturally satisfied positivity \cite{laan_targeted_2011, laan_targeted_2018}. We defined our research question using an MTP, which aims to quantify the impact on the counterfactual outcome of an intervention that shifts the exposure for each observation to a new level $A = a^d$, set by a deterministic function of the observed `natural' exposure level $A = a$ and covariates $W = w$ through some shift function $d(A,W)$ \cite{haneuse_estimation_2013}. For example, we might consider an MTP where the mobility shift amount depended on population density: sparse counties would be assigned a mobility reduction of 10\% while dense counties would be assigned 20\%. However, for simplicity and demonstration, we ignored covariates when determining shift amounts, instead examining the expected counterfactual outcome if the mobility distribution across counties were shifted by an additive or multiplicative constant - say, if the percentage of people staying in their home all day were increased by 5 points from the observed level in each county.

The MTP approach was attractive in our context because it allows, as per Haneuse and Rotnitzky \cite{haneuse_estimation_2013}, ``the set of feasible treatments for each subject [to depend] on attributes that are not fully captured by baseline covariates but which are reasonably captured by the treatment actually received." There is no assumption that, for example, the densely populated New York County would be shifted to the mobility level of a rural area, even if some observed covariate values (say, demographics) were similar.  Full details of the assumptions and estimation strategy are given in Section \ref{section:SciQandEstimand}. MTP is similar to a \emph{stochastic intervention}, where the exposure $a^d$ is a random draw from a shifted distribution $A^d$ of the observed exposures $A$ \cite{didelez_direct_2012, munoz_population_2012, stock_nonparametric_1989}, possibly depending on covariate values. The MTP approach has been used in several biomedical settings  \cite{diaz_causal_2020, kamel_relationship_2019, hubbard_time-dependent_2013} and in at least one public policy context \cite{rudolph_when_2021}, but to our knowledge this is the first time it has been applied to infectious disease.

\section{Methods}
\label{section:SciQandEstimand}

At the county-level, we aimed to evaluate the effect of shifting mobility on new COVID-19 cases per 100,000 residents two weeks later during the period from June 1 - November 14, 2020. Data on confirmed COVID-19 cases were obtained from the  New York Times \cite{the_new_york_times_coronavirus_2020} and USAFacts \cite{noauthor_coronavirus_nodate}. In line with other studies \cite{badr_association_2020, li_association_2020}, a two-week ``leading'' period was selected based on the reported incubation time for COVID-19 \cite{linton_incubation_2020, lauer_incubation_2020, li_early_2020} and to account for delays in case reporting. In sensitivity analyses, we examined the impact on  one-week leading case rates. 

We considered 10 mobility indices, generated from mobile phones and selected to represent a range of behaviors that potentially impact and are impacted by COVID-19 case rates. Specifically, Google's ``residential'' index \cite{google_covid19_nodate} and Facebook's ``single tile user" and ``tiles visited" indices \cite{facebook_movement_nodate} were expected to capture behavior related to sheltering at place at home. Additional indices from Google \cite{google_covid19_nodate} captured changes in workplace attendance (``workplaces''), non-essential travel (``retail and recreation"), and use of public transportation (``transit stations''). Likewise, the ``Device Exposure Index" (DEX) and ``Adjusted Device Exposure Index" (DEX-A) \cite{couture_measuring_2020} captured the density of human traffic at locations visited. Finally, Descartes Labs' ``m50" and ``m50 Index" \cite{descartes_labs_mobility_nodate} captured distance traveled outside the home (full index definitions in Appendix \ref{appendix:exposures}). The ten indices were only reported in counties where sufficient observations were available to provide accurate and anonymous data, typically more populous counties. To address this and make the results more comparable across different indices, we excluded counties with fewer than 40,000 residents. This led to the exclusion of 1,948 of 3,139 counties (62\%) in our main analysis, many of them rural and sparsely populated. The 1,182 included counties represent 90\% of the U.S. population. A comparison of the counties included/excluded is in Appendix \ref{appendix:counties}. Notably, for interpretation of results, this inclusion/exclusion criteria changes our target population from all to just the most populous U.S. counties.

To account for measured differences between counties on factors potentially influencing mobility and case rates (i.e., confounders), we collected a wide range of covariates from public sources. These spanned several domains: socioeconomic (ex: levels of poverty, unemployment, education, median household income, crowded housing, population density, urbanization); health (ex: rates of smoking and obesity, pollution, extreme temperatures); demographic (ex: distribution of age, race, sex); and political (ex: 2016 presidential election result, mask mandate level). Additionally, since the one-week-behind case rate could influence that week's mobility (e.g., people may be less mobile if the perceived danger is higher) and subsequent transmission, previous week case rates were also included in the confounder set $W$. The full list of variables collected and links to their sources are provided in Appendix \ref{appendix:covariates}.

\subsection{Data structure, causal model, \& causal parameter} 
 
To assess impact over time while avoiding holiday- and weekday-level case reporting idiosyncrasies \cite{noauthor_analysis_nodate}, mobility and case data were binned into weeks using a simple mean. We then repeated our analysis in cross-sectional snapshots over each week during the period beginning June 1, 2020 and ending November 14, 2020 and separately for each mobility index. Our observed data consisted of our outcome $Y$, the new case rate; exposure $A$, the mobility level; and the set of potential confounders $W$. For each week and mobility index, we assumed that the following nonparametric structural equation model (NPSEM) characterized the data generating process \cite{pearl_causality_2000}: 

\begin{align*}
  W &= f_W(U_W)\\
  A &= f_A(W, U_A) \\ 
  Y &= f_Y(W, A, U_Y)
\end{align*}

where $U_W$, $U_A$, and $U_Y$ are unobserved random variables and $f_W, f_A$, and $f_Y$ are unspecified (non-parametric) functions. This causal model also encoded our assumption of independence between counties. (We discuss the plausibility of this assumption below.) For each week and index, we assumed the county-level observed data $O=(W,A,Y)$ were generated by sampling from a distribution $\mathbb{P}_0$ compatible with this causal model.

\begin{figure}[h]
\begin{center}
\includegraphics[width=\textwidth]{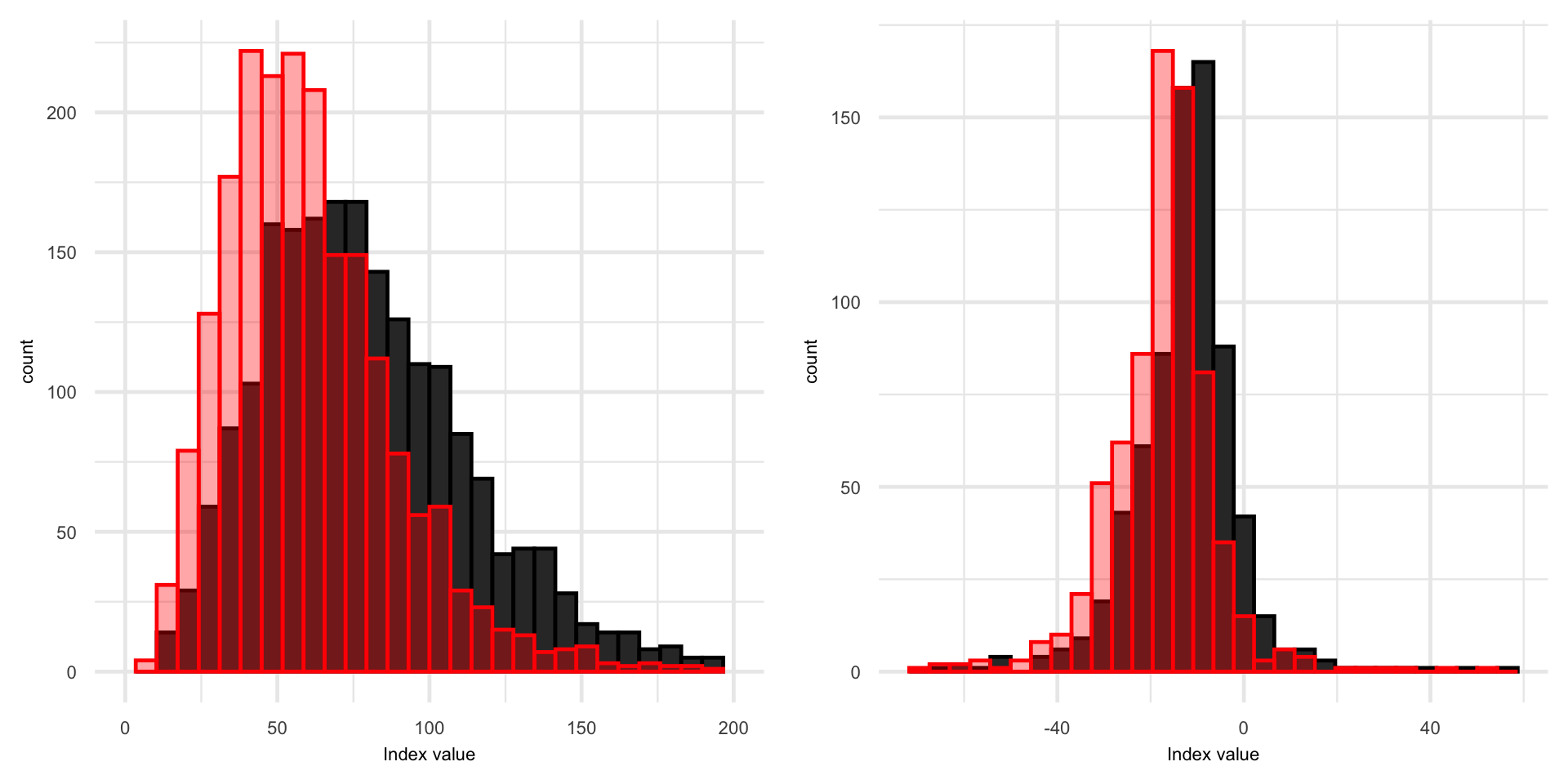}
  \caption{Example distribution shifts exposures. Observed values black, shifted values red.}
    \label{fig:shift}
\end{center}
\end{figure}

Following \cite{diaz_nonparametric_2021}, we then modified $Y = f_Y(W, A^d, U_Y)$ in the causal model to generate counterfactual case rates under the shifted exposure distributions. In our application, shifts did not depend on the covariates $W$, so simplified the treatment rule to $d(A;M) = MA$ and $d(A;C) = C + A$ for shifts by a multiplicative constant $M$ or additive constant $C$, respectively. We selected shift amounts to reduce mobility based on each index definition and range of observed values. For example, Facebook's ``single tile user'' is a proportion of users who stayed in a single location (presumably, their residence) in a day and is thus bounded by 0 and 1. Based on its observed distribution (pooling over all follow-up weeks), we considered relative increases of 5, 7, and 8 percentage points, truncated at the upper bound of 1. For some indices, such as the DEX-A, which measures the density of mobile devices visiting commercial locations, larger multiplicative shifts were reasonable; an example of a reduction shift of 0.75 is shown in the left panel of Figure \ref{fig:shift}. As a third example, consider Google's ``retail and recreation" index, defined as the percentage of visits to non-essential businesses relative to a pre-COVID baseline of 100\%. Here, we examined additive shifts of -4, -5, and -6 percentage points, bounded by the index limit of -100, shown in the right panel of Figure \ref{fig:shift}. A list of shifts examined for each variable is given in Appendix \ref{appendix:exposures}.

Given the selected shift $A^d$ for each mobility index, we then specified the causal parameter as the expected counterfactual COVID-19 case rate if the mobility index had been shifted to $A^d$ versus if mobility index remained at its observed level $A$: 

\begin{equation}
    \psi = \mathbb{E}\left[ Y(A^d) \right]- \mathbb{E}[Y(A)],
\label{Eq:Causal}    
\end{equation}

with $d(A;M) = MA$ for multiplicative shifts or $d(A;C) = C + A$ for additive shifts. Setting $M=1$ or $C=0$ in the treatment rule $d$ recovers the observed exposure $A$ and outcome $Y$.

\subsection{Identification \& the statistical estimand}
\label{section:timeidentification}

For the causal parameter (Eq.~\ref{Eq:Causal}, written in terms of a summary measure of the distribution of counterfactuals) to be identified in terms of the observed data distribution, several assumptions were required, discussed in more detail in Appendix \ref{appendix:assumptions} \cite{munoz_population_2012, haneuse_estimation_2013, diaz_nonparametric_2021}:

\begin{itemize}
    \item No unmeasured confounding: There are no unmeasured common causes of the mobility index and subsequent case rates, which we can formalize as $Y(A^d) \perp\!\!\!\perp A | W$. In our context, this assumption would be violated if, for example, an unmeasured variable influenced both the county-level mobility and the case rate. One possible example could be issuing of and compliance with NPIs; an NPI might reduce mobility and also influence non-mobility-related behavior that is associated with future case rates. We attempted to improve the plausibility of this assumption by considering a very large confounder set including known correlates for our outcome and exposure. Nevertheless, we cannot guarantee that this assumption holds, and therefore limit our interpretations to statistical associations rather than causal effects.
    
    \item Positivity: If $(a, w)$ is within the support of $A, W$, then $(d(a, w), w)$ must also be within the support of $A, W$. In practice, this means that for any given time-slice and set of adjustment covariates, there is a positive probability of finding a county with the same covariate values and a mobility level matching the shifted value.
    
    \item Independence of counties, consistency, and time-ordering.
\end{itemize}

If these identifiability assumptions held, we could specify and focus our estimation efforts on a statistical estimand that equals the wished-for causal effect. In the (likely) case that they are not satisfied, we could still specify and focus our estimation efforts on a statistical estimand that is as close as possible to the causal parameter. Factoring the joint distribution of the observed data $P_0$ into $P_0(O) = P_0(Y|A,W)P_0(A|W)P_0(W)$, it was shown in \cite{haneuse_estimation_2013} that the statistical estimand corresponding to expected counterfactual outcome under treatment rule $d$, $\mathbb{E}[Y(A^d)]$, is given by  

\begin{equation}
\label{robins}
\begin{aligned}
   \psi_0(A^d) = \int \mathbb{E}(Y \mid A=a^d,W) \text{d}F_{A,W}(a,w),
\end{aligned}
\end{equation}

with $\text{d}F_{A,W}(a,w)$ the joint density of received exposures $A$ and covariate levels $W$ being integrated over, equivalent to the extended g-computation formula of Robins et al. \cite{robins_effects_2004}. Extending this, our statistical estimand of interest, corresponding to the expected difference in COVID-19 case rates under shifted and observed mobility, is given by $\psi_{\Delta}$:

$$
\psi_{\Delta} = \psi_0(A^d) - \psi_0(A) = \int \mathbb{E}(Y \mid A=a^d,W) \text{d}F_{A,W}(a,w) - \mathbb{E}_{P_0}(Y),
$$

where $Y$ denotes the new case rate and $P_0$ the observed data distribution. In words, our statistical estimand $\psi_{\Delta}$ is the difference in expectation of new cases per 100,000-residents under a specified mobility shift $A^d$, after adjusting for measured confounders, and the observed expectation of new cases per 100,000-residents $\mathbb{E}_{P_0}(Y)$. All interventions we examined were reductions in mobility; hence, under this definition, we expect $\psi_{\Delta}$ to be negative if mobility reductions are associated with lower case rates.

\section{Estimation}
\label{section:Estimation}

To estimate the expected outcome under the observed exposure $\psi_0(A)$, we used the empirical mean outcome. For estimating the shifted parameter $\psi_0(A^d)$, we used TMLE \cite{laan_targeted_2011, laan_targeted_2018}. TMLE is a general framework for constructing asymptotically linear estimators with an optimal bias-variance tradeoff for a target parameter; in our case, $\psi_0({A^d})$. Typically, TMLE uses the factorization of $P_0(O)$ into an outcome regression $\bar{Q}(A,W) = \mathbb{E}_{P_0}(Y \mid A,W)$ and an intervention mechanism $g(A \mid W) = P(A \mid W)$, following which an initial estimate of the outcome regression $\bar{Q}$ is updated by a fluctuation that is a function of the intervention mechanism. The update step tilts the preliminary estimates toward a solution of the efficient influence function estimating equation, endowing it with desirable asymptotic properties such as double-robustness, described in Appendix \ref{appendix:estimation}. Further, it allows practictioners to utilize machine learning algorithms in the estimation of $\bar{Q}$ and $g$ while still maintaining valid statistical inference.

TMLE is a general framework; the exact method may vary depending on the data structure and question of interest. In this application, we used the implementation of Díaz \cite{diaz_nonparametric_2021}. We summarize briefly the algorithm; further technical details are in \cite{diaz_nonparametric_2021}.

First we define the density ratio

\begin{equation}
\begin{aligned}
    r(a, w) &= \frac{g^d(a|w)}{g(a|w)} \\
\end{aligned}    
\end{equation}

where $g^d(a|w)$ is the conditional density of $a$ under the shifted distribution and $g(a|w)$ the conditional density as observed. Unfortunately, estimating the conditional density $g$ directly can be difficult in high dimensions. Few data-adaptive algorithms exist \cite{diaz_nonparametric_2021}, and for those that do \cite{benkeser_highly_2016}, computation time can be demanding and data sparsity can make estimates unstable. Luckily, an equivalent estimate for $r$ can be found by recasting it as a classification problem \cite{diaz_nonparametric_2021}. In this formulation, a data set consisting of two copies of each observation is created, one assigned the observed exposure and one the shifted exposure. Further, each is given an indicator $\Lambda$ for whether it received the natural or shifted value of treatment. Using Bayes' rule and the fact that $P(\Lambda = 1) = P(\Lambda = 0)= .5$,

\begin{equation}
\begin{aligned}
    r(a, w) = \frac{g^d(a|w)}{g(a|w)} &= \frac{P(a,w \mid \Lambda = 1)}{P(a,w \mid \Lambda = 0)} = \frac{P(\Lambda = 1 \mid a, w)}{P( \Lambda = 0 \mid a, w)} \\
    &=\frac{P(\Lambda = 1 \mid a, w)}{1 - P( \Lambda = 1 \mid a, w)}.
\end{aligned}    
\end{equation}

In this formulation, the density ratio estimate can be calculated as a function of $P(\Lambda = 1 \mid a, w)$, which can be estimated using any binary classification method and which we will write as $\Tilde{\lambda}$. This innovation by Díaz \cite{diaz_nonparametric_2021} drastically improved computation time compared to estimating $g$ directly (as is done in several algorithms for stochastic interventions \cite{hejazi_txshift_2020, hejazi_tmle3shift_2021}).

To implement estimation of $\Tilde{\lambda}$ and the conditional outcome regression $\bar{Q}$ flexibly, we used the ensemble machine learning algorithm Super Learner, allowing us to minimize parametric assumptions \cite{laan_super_2007}. Super Learner creates a convex combination of candidate algorithms chosen through cross-validation to minimize a user-specified loss function. In our case, the Super Learner ensemble included generalized linear models, generalized additive models, gradient boosted decision trees, random forest, and a simple mean.

With nuisance function estimates in hand, step-by-step estimation of and inference for $\psi_0(A^d)$ is outlined in Appendix \ref{appendix:estimation}. All statistical analysis was performed in R version 4.0.3 \cite{r_core_team_r_2020}, in particular the \verb|ltmp_tmle| function in the \verb|lmtp| package \cite{williams_lmtp_2020}. Code and data for reproduction is available at the authors' GitHub repository. 

\section{Results}
\label{section:Results}

The following figures show estimates of $\hat{\psi}_{\Delta}$, the difference between the COVID-19 case rate that would be expected under the shifted distribution, after adjusting for measured confounders, and the observed average case rate, with associated 95\% confidence intervals (CIs), for weekly time slices from June 1, 2020 to November 14, 2020. For comparison, we also show unadjusted estimates, which is defined analogously to $\hat{\psi}_{\Delta}$, simply without any covariates $W$.

\begin{figure}[!h]
\begin{center}
\includegraphics[width=.5\textwidth]{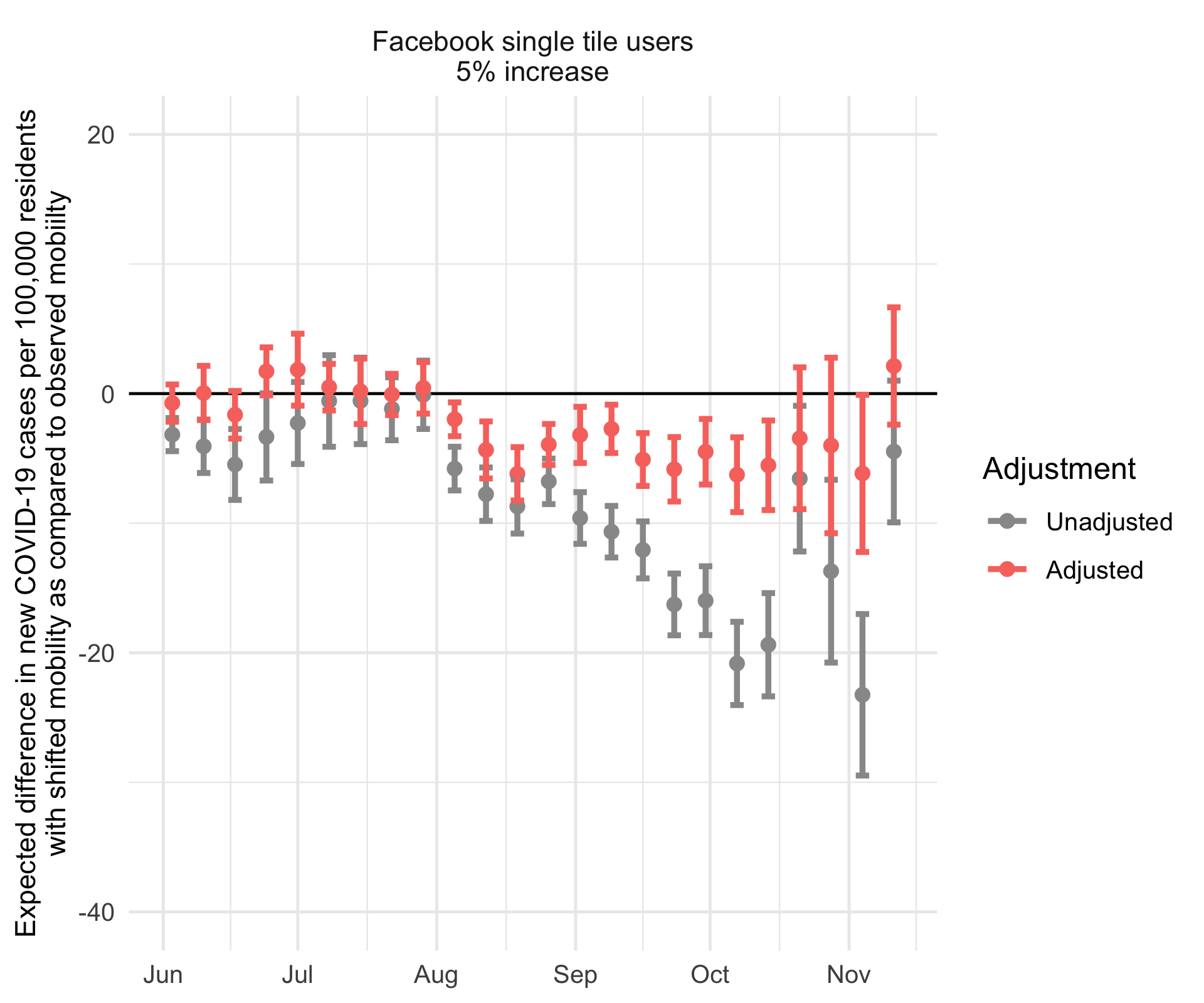}
  \caption{Difference in mean predicted case rates two weeks ahead with shifted distribution vs. observed case rates for weekly time slices from June 1 - November 14, 2020. Unadjusted analyses show some correlation between mobility and case rates (negative gray values), but after adjustment (colored values), there is no consistent association across the set of weeks examined. The +5\% increase corresponds to a reduction in mobility (more people staying at home).}
    \label{fig:example}
\end{center}
\end{figure}

As an example, consider Figure \ref{fig:example}, showing $\hat{\psi}_{\Delta}$ for ``Facebook single tile users," which measures the proportion of mobile devices in a county that stayed in a single location (presumably, a home) each day, averaged over the week. In the first time-slice (June 1, 2020), the unadjusted results, in gray, showed a small association between reduced mobility (5\% increase in the proportion of people staying at home) and reduced case rates: -3.2 (95\% CI: -4.4, -1.9) cases per 100,000 residents two weeks later. After adjusting for confounders (in red), the change in case rates shrinks to -0.7 (-2.2, 0.7). In the week of September 28, the results are similar but more extreme: a 5\% increase in the proportion of people staying at home in the unadjusted analysis suggests a change of -16.0 (-18.6, -13.3) cases per 100,000 residents two weeks later, while that association is attenuated to -4.5 (-7.0, -2.0) after confounder adjustment. Across all of the time slices for that mobility index, we see the same pattern: Unadjusted estimates of $\hat{\psi}_{\Delta}$ vary, but are negative, as we would expect; lower mobility appears to be associated with lower case rates. After adjustment, however, the estimates shrink towards zero and many of the CIs overlap with zero.

\begin{figure}[!h]
\begin{center}
\includegraphics[width=\textwidth]{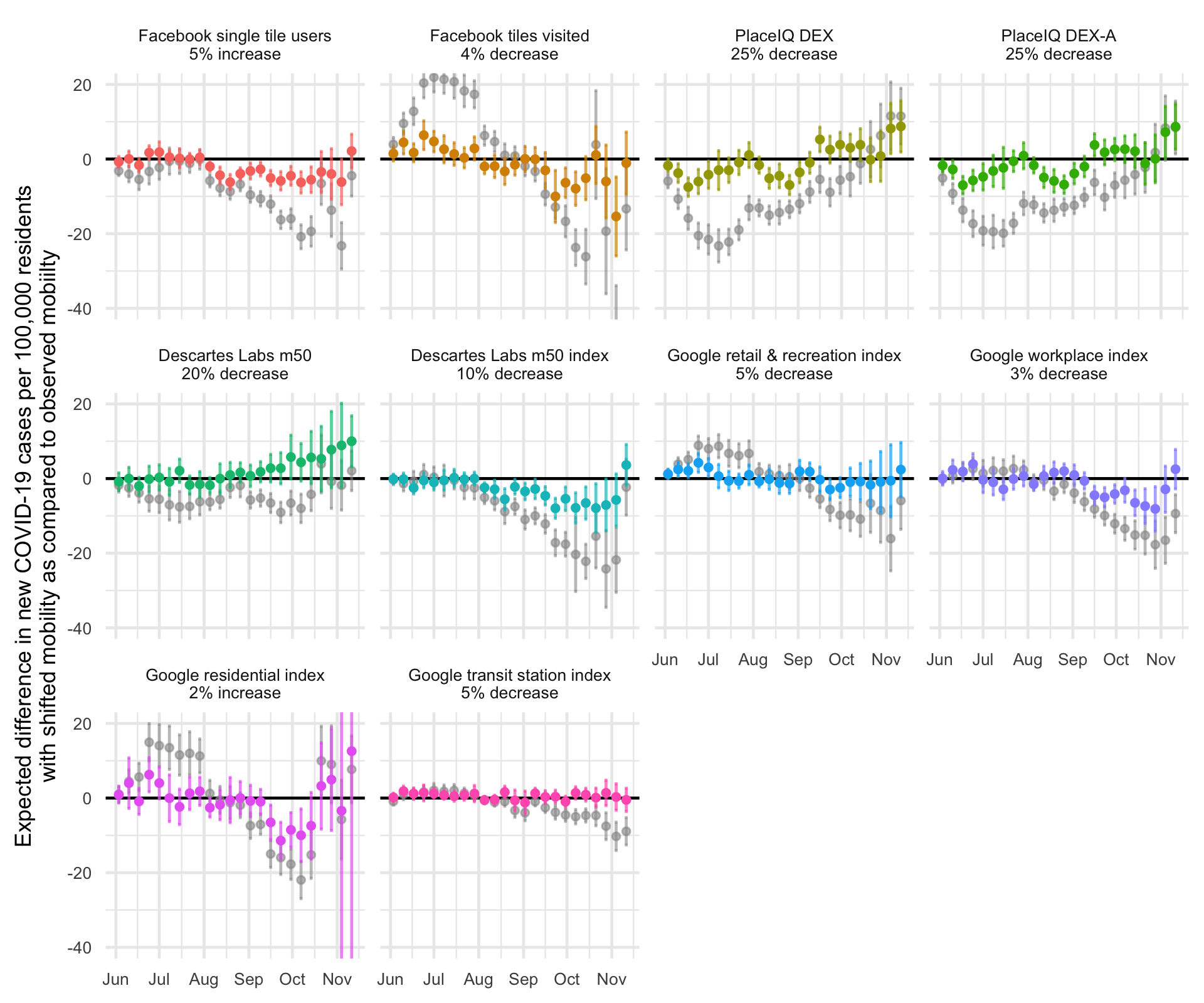}
  \caption{Difference in mean predicted case rates two weeks ahead with shifted distribution vs. observed case rates for weekly time slices from June 1 - November 14, 2020. Unadjusted analyses show some correlation between mobility and case rates (gray values with 95\% CIs below the null line of 0), but after adjustment (colored values), no indices show consistent associations across the weeks examined. For most indices, the majority of confidence intervals include 0 after adjustment. Recall that the two stay-at-home measures, Facebook's `single tile users' and Google's  `residential', were assigned positive shifts, which corresponded to a reduction in mobility.}
    \label{fig:MAIN}
\end{center}
\end{figure}

Broadening the scope to all indices in Figure \ref{fig:MAIN}, some patterns emerge. First, almost all unadjusted estimates are larger than the adjusted ones, though point estimates are usually in the same direction after adjustment. Second, CIs increase in October and November, suggesting higher outcome variance, more treatment effect variability, or higher density ratios (perhaps from near practical positivity violations). Third and most notable, after adjustment for our confounder set, associations were generally small, CIs overlapped with zero, and point estimates showed almost no consistent trends of positive or negative association.

For example, only six of the indices showed even short-term consistent associations with the outcome after adjustment, perhaps due to seasonal effects: Facebook's `single tile users' metric (Aug, Sept, Oct), PlaceIQ's `DEX' and `DEX-A' measures of device density (Jun, Aug), Decartes Labs `m50 index' of distance traveled from the home (Aug, Sept, Oct), Google's workplace index (Sept, Oct), and Google's stay-at-home residential index (Sept, Oct). However, these same indices had uncertain associations in other time periods, with CIs overlapping zero, and in several instances had point estimates (and sometimes and CIs) unexpectedly above zero.

Surprisingly, three indices showed positive unadjusted point estimates in June and July (Facebook `tiles visited' and Google's `retail \& recreation' and `residential' indices), with Facebook's `tiles visited' index showing  an unusually large swing between summer and fall. After adjustment, the estimates were strongly attenuated. These data points may be displaying most clearly the confounding effect of the covariates in our adjustment set. More discussion of possible explanations for the mixed results are presented in Section \ref{section:Discussion}. We performed several sensitivity analyses that did not meaningfully change our results (Appendix \ref{appendix:sensitivity}).

\section{Discussion}
\label{section:Discussion}

To the best of our knowledge, we are the first to apply a doubly-robust MTP/TMLE approach to provide insights into a pressing question in infectious disease research. Specifically, we aimed to answer: Did reductions in mobility, as captured the mobile phone data, lead to reduction in subsequent COVID-19 case rates across U.S. counties in the summer and fall of 2020? We also hope this paper will serve as a demonstration of how MTP can be used by other researchers to minimize (likely unjustifiable) parametric assumptions when defining causal effects and how TMLE can be applied to estimate the corresponding statistical parameters. We again note that the causal inference literature for binary exposures is vast and well-established, especially as compared to the corresponding literature for continuous exposures. We believe MTP is a powerful and under-utilized approach to defining effects for continuous exposures.

Our findings on how the associations between mobility and COVID spread vary over time and with confounder adjustment are in line with the results found in the earlier research outlined in Appendix \ref{appendix:lit}. Though our unadjusted analyses seem to suggest that decreased mobility was associated with reductions in COVID-19 case rates for the majority of timepoints examined, adjusted results suggest that the associations between commonly studied indices and COVID-19 case rates may be highly confounded by county characteristics and current case rates. For example, our adjustment set included variables for county-level political lean, which may be a proxy for adherence to NPIs like mask-wearing mandates. It could appear that counties with lower mobility have fewer new cases, but if low-mobility counties are also more consistently adhering to mask-wearing due to underlying political dynamics, the association with mobility would vanish after adjustment. Including one-week lagged case rates in the adjustment set (as we argue is necessary in the infectious disease context) also demonstrates important confounding by epidemic arrival time and intensity.

There are several limitations to our analysis, some of which may explain the inconsistent patterns of association we see over time and suggest avenues for future investigation. First, while we chose a timeframe when COVID testing was widely available, differential ascertainment of case rates across counties could bias our estimates. Second, while county-level mobility and covariate data are easily accessible, there might be important sub-county level effects that are masked when aggregating to the county level. Performing this analysis at the zip code or census tract level might reveal stronger or more consistent associations. Indeed, an analysis of zip-code level mobility data and case rates in five cities \cite{glaeser_how_2020}, while finding a strong association, also reported substantial heterogeneity across space and time. Third, our assumption of independence between counties is unlikely to hold, given state-level policy enactment and spillover effects between counties. Network effects can be unpredictable \cite{lee_network_2021}, so we hesitate to speculate about the directions in which they might have biased our point estimates or standard errors, but they are likely in play.

Fourth, despite our large adjustment set, it is possible that we excluded important confounders, or that we have unmeasured confounding, as is common with observational data. For example, we had no data on the strength of a county's public health infrastructure or compliance with NPIs, which could presumably impact both the mobility and case rates. Related to the dimension of the adjustment set $W$, practical positivity violations may have also limited our ability to detect associations. After adjusting for demographics, political lean, and economic variables, the variability in exposure and outcome across counties may have been small, limiting the amount of shift we could apply without generating extreme density ratios and the accompanying large standard errors, since the density ratio is a key component of the efficiency bound. If mobility index data were available for the sparsely populated counties that we excluded, it might be possible to expand our target population and investigate a wider range of intervention shift values that might detect stronger or more consistent associations. Fifth, there are still many unknowns about the dynamics of COVID transmission over space and time. Including one week lagged case rates in our adjustment set may have helped address some of those unknowns, but is likely not able to account for all of the complex dynamics of COVID transmission. For example, the return to schools in fall 2020 may partially explain how several indices show different behavior in the summer and fall: the adjusted DEX and DEX-A point estimates change sign and the adjusted m50 estimates shift from minimal associations to stronger ones. Finally, our choice of a cross-sectional time-series analysis in weekly slices may have hidden complex patterns over time. A longitudinal analysis that examines the full scope of mobility over time and its impact on cumulative cases per capita could provide a more complete view of the full impact (or lack thereof) of mobility on case transmission. Doubly-robust longitudinal MTP methods are available in the \verb|ltmp| R package and offer a promising avenue to answer those questions.

Despite these limitations, our analysis provides evidence that mobility indices may not be a simple proxy measure for future case rates with airborne infectious diseases. Using MTP allowed us to consider shifts in mobility supported by our observed data, and TMLE allowed for the leveraging of machine learning to minimize parametric assumptions while still providing statistical inference. For continuous exposures, MTP can be a useful tool for important research and policy questions.

\printbibliography

\clearpage
\newpage

\begin{appendices}

\section{Brief review of the literature on mobility \& COVID-19 transmission}
\label{appendix:lit}

Recent papers investigating the associations between mobility, COVID-19 spread, and NPIs can be roughly categorized into one of three bins: work that uses mobility as an input into a more complex model to forecast COVID-19 cases and deaths (see, e.g., \cite{chang_mobility_2020, chernozhukov_causal_2021, karaivanov_face_2020}); work that assesses the effectiveness of, or compliance with, NPIs, using mobility as an outcome variable (see, e.g., \cite{jay_neighbourhood_2020, gupta_tracking_2020, lai_unsupervised_2020, yabe_non-compulsory_2020, allcott_what_2020, wellenius_impacts_2021}); and finally work that treats mobility as an exposure variable and new COVID-19 case rates or deaths as the outcome. This work falls clearly in the last category.

Most papers focusing on the early-pandemic (Spring 2020) timeframe found strong correlations between mobility and COVID-19 spread \cite{li_association_2020, badr_association_2020, xiong_mobile_2020, persson_monitoring_2021, wellenius_impacts_2021}. However, these early studies are potentially confounded by the spatial dynamics of where the pandemic arrived first and unmeasured variables such as underlying changes in behavior including mask-wearing or the propensity to socialize outdoors rather than in closed indoor environments (see, e.g., the heterogeneity identified by \cite{allcott_what_2020}). In addition, they may be subject to bias from undercounted cases due to lack of testing capacity in many parts of the U.S. \cite{noauthor_covid_nodate}. Given these challenges, it is perhaps unsurprising that the associations between mobility and COVID-19 cases were strongly attenuated in studies extended through mid-Summer of 2020 \cite{badr_limitations_2020, gatalo_associations_2020}.

Our findings on how the associations between mobility and COVID spread vary over time and with confounder adjustment are in line with other results. In an analysis using data from the 25 hardest-hit U.S. counties through mid-April 2020, Badr et al. \cite{badr_association_2020} fit an unadjusted statistical model for future case growth rates based on a mobility measure using number of trips made per day compared to a pre-pandemic baseline. While strong associations were found between changes in their mobility index and county-level case growth 9-12 days later, a later updated analysis \cite{badr_limitations_2020}, using data through September 16 and expanding the dataset to all U.S. counties, showed that the association disappeared after April. Similarly, Gatalo et al. \cite{gatalo_associations_2020}, using a slightly different methodology, found strong correlations between mobility and case growth from March 27 to April 20, but only a weak correlation from then until July 22. Another analysis \cite{xiong_mobile_2020}, using data through June 9, 2020, treated the inflow of traffic into a county as the exposure, adjusted for a small number of socio-demographic covariates, and found a strong relationship between inflow and new cases 7 days later. Finally, an examination \cite{li_association_2020} of the correlations between Google's six mobility indices (retail \& recreation, transit stations, workplaces, residential, parks, grocery \& pharmacy) and 11-day ahead case growth rates at the county level, stratifying by urban-rural classification, suggested strong associations in the February to April timeframe, especially in more dense counties, but those correlations were weaker when the timeframe was extended to June.

Our work adds to other evidence of confounding in COVID-19 research; for example, McLaren \cite{mclaren_racial_2020} found that some correlations between racial/ethnic groups and COVID-19 mortality disappeared once confounding factors were taken into account, and Wong and Balzer \cite{wong_evaluating_2021} found that associations between mask mandates and future case rates were attenuated after confounder adjustment. We suspect this confounding may explain why our results contradict those of Wellenius et al. \cite{wellenius_impacts_2021}, who found that mobility levels (using many of the same county-level metrics) in Spring 2020 were strongly associated with COVID case growth rates two weeks later. Their work differed from this analysis in that the researchers did not include any socioeconomic covariates in their parametric model nor extend their analysis beyond March and early April.

\section{Data details}
\label{appendix:data}

\subsection{Mobility index sources, definitions, and shifts considered}
\label{appendix:exposures}

\textbf{Facebook Movement Range}
\begin{itemize}
    \item Change in movement: Percentage change in the number of `Bing tiles' (roughly 600 m by 600 m) a device pinged in a given day, compared to the baseline average for that weekday in February 2020, excluding Presidents' Day, aggregated at the county level. We examined additive shifts of -3 and -4.
    \item Stay put: Percentage of people who stay in a single Bing tile over an entire day. We examined multiplicative shifts of 1.05, 1.07, and 1.08.
\end{itemize}

\textbf{Google COVID-19 Community Mobility Reports}

For Google indices, all of the baseline values are defined as the median value for that day of the week from January 3, 2020, to February 6, 2020.

\begin{itemize}
    \item Retail \& recreation: Percentage change in number of visits and length of stay from baseline for restaurants, malls, movie theaters, and retail stores (excluding grocery stores and pharmacies). We examined additive shifts of -4, -5, and -6.
    \item Transit stations: Percentage change in number of visits and length of stay from baseline for mass transit stations (buses, subways, etc). We examined additive shifts of -5, -10, and -15.
    \item Workplaces: Percentage change in number of visits and length of stay from baseline for an individual's place of work. We examined additive shifts of -2, -3, and -4.
    \item Residential: Percentage change from baseline of duration time spent at one's residence each day. Since the baseline values are high, the variation in this index is smaller than the others. We examined additive shifts of +1, +2, and +3.
\end{itemize}

\textbf{PlaceIQ DEX}
\begin{itemize}
    \item Device exposure index (DEX): number of distinct devices that entered the same commercial spaces that day as the target device, then averages that number across all devices in the county. We examined multiplicative shifts of 0.75, 0.7, and 0.65.
    \item Adjusted DEX: Scaled version of the DEX that adjusts for the possibility of devices sheltering in place, generating no pings, and hence not having their zero-DEX values be counted in the overall average. We examined multiplicative shifts of 0.75, 0.7, and 0.65.
\end{itemize}

\textbf{Descartes Labs m50}
\begin{itemize}
    \item m50: Median maximum distance traveled from starting location by all tracked devices in a county. We examined multiplicative shifts of 0.6, 0.7, and 0.8.
    \item m50 index: Relative value of m50 relative to median baseline value for that day of the week between February 2, 2020, and March 7, 2020. We examined multiplicative shifts of 0.9 and 0.85.
\end{itemize}

\subsection{Comparison of counties included/excluded}
\label{appendix:counties}

All mobility indices require sufficient sample size within a county to report reliable data on a given day. In low-population counties, this can make the mobility data sparse or nonexistent over the timeframe in question. Hence, to improve comparability across different indices and address positivity violations, we restricted our analysis to counties with 40,000 or more residents. The figure below compares the covariate distribution across the included/excluded counties. While a vast number of counties were excluded, they represent only 10\% of the U.S. population, and covariate means were similar to the counties included.

\begin{center}
\begin{tabular}{l|ccc } 
 \hline
 & All counties & Included & Excluded \\
 \hline
N & 3,158 & 1,182 & 1,926 \\
Mean population & 105,362 & 250,886 & 16,053 \\
Mean population density (/sq. mi) & 27,167 & 62,366 & 5,608 \\
\% Adults without health insurance & 14.3\% & 13\% &  15\% \\  
\% Under 18 & 22.3\% & 22.6\%& 22.2\% \\
\% Households below poverty line & 15.2\% & 13.8\% & 16.1\% \\  
\% Hispanic & 9.3\% & 10.8\% & 8.4\% \\
\% Women & 49.9\% & 50.6\% & 49.5\% \\
\% Black &  8.9\% &  9.9\% & 8.5\% \\
\% Clinton vote in 2016 election & 31.7\% & 38.6\% & 27.8\% \\
\% Drive alone to work & 79.2\% &  80.5\% &  78.9\% \\
 \hline
\end{tabular}
\end{center}

\subsection{Adjustment variables and their sources}
\label{appendix:covariates}

We gathered a rich set of social, economic, physical, and demographic variables from the U.S. Census American Community Survey, U.S. Bureau of Labor Statistics, U.S. Centers for Disease Control, U.S. Department of Agriculture, University of Washington Population Health Institute and Department of Political Science, National Oceanic and Atmospheric Administration, and MIT Election Data and Science Lab. These included the covariates below, which we screened into the adjustment set after examining their correlations with the exposure (mobility index) and outcome (leading new cases per 100,000 people).

\begin{itemize}
    \item U.S. Census American Community Survey
    \begin{itemize}
        \item Percentage of households below the poverty line
    \end{itemize}
    \item University of Washington Population Health Institute, via Robert Wood Johnson Foundation County Health Rankings
    \begin{itemize}
        \item Percentage of people that identify as Black
        \item Percentage of people that identify as Hispanic
        \item Percentage of people that drive alone to work
        \item Percentage of people that are women (note: values under 49\% are often an indicator of a large prison being in the county)
        \item Percentage of people that are under 18 years of age
        \item Percentage of adults without health insurance
    \end{itemize}
    \item MIT Election Data and Science Lab
    \begin{itemize}
        \item Percentage of vote for Hillary Clinton in the 2016 Presidential election
    \end{itemize}
\end{itemize}

The following covariates were also examined, but were removed from the final adjustment set after the screening process.

\begin{itemize}
    \item U.S. Census American Community Survey
    \begin{itemize}
        \item Population density
        \item Natural logarithm of population density
        \item Average household size
        \item Percentage of people living below the poverty line
        \item Percentage of households with income below the poverty line
        \item Percentage of people in the county who live outside of a metropolitan area
    \end{itemize}
        \item University of Washington Population Health Institute, via Robert Wood Johnson Foundation County Health Rankings
    \begin{itemize}
        \item Percentage of adults who smoke
        \item Percentage of people that identify as Asian
        \item Percentage of people that are 65 years of age or older
        \item Percentage of people with more than a high school education
        \item Index of income inequality
        \item Natural logarithm of median household income
        \item Level of air particulate matter (pollution)
    \end{itemize}
    \item State-level social distancing policies in response to the 2019 novel coronavirus in the U.S., University of Washington Department of Political Science
    \begin{itemize}
        \item Mask mandate level (statewide data only)
    \end{itemize}
    \item Centers for Disease Control and Prevention
    \begin{itemize}
        \item Percentage of people who live in crowded housing
        \item Percentage of people categorized as obese
        \item Percentage of people living with diabetes
    \end{itemize}
    \item National Oceanic and Atmospheric Administration
    \begin{itemize}
        \item Absolute daily high temperature deviation from 70 degrees Farenheit
    \end{itemize}
    \item U.S. Bureau of Labor Statistics
    \begin{itemize}
        \item Unemployment rate
    \end{itemize}
    \item U.S. Department of Agriculture
    \begin{itemize}
        \item Presence of a level 5 meatpacking plant (largest plants)
        \item Presence of a level 4 meatpacking plant (second largest plants)
        \item Presence of a level 4 or level 5 meatpacking plant
    \end{itemize}
\end{itemize}

\section{Details of identification assumptions}
\label{appendix:assumptions}

The positivity assumption in this approach is a weaker positivity assumption than for other causal effects, which require nonzero probability of all observations receiving a specified level of treatment (say, ``low" or ``high" mobility) within values of adjustment variables. While reasonable in theory, this is viable in practice only when the specified shifts in $A$ are not too drastic and the adjustment set $W$ not too large. Our main analysis restricted to shifts that were small enough to make this assumption tenable, while recognizing that this approach makes our causal effect data-adaptive.

Given plausible concerns about data support and desire to avoid controlling for instrumental variables, we reduced the adjustment set $W$ to eight variables based on which had the strongest univariate associations with the outcome and exposure, pooled over all time slices. In supplementary analyses, we also tested strategies where (i) covariates were chosen week-by-week instead of being held constant across all time slices and (ii) a smaller covariate set was chosen to make possible examinations of larger shift values without practical positivity violations.

Beyond no unmeasured confounding and positivity, there are three additional assumptions that are implied by our adoption of the NPSEM. (1) Independence of counties. This independence assumption is likely violated since there were state-level policies (e.g., stay-at-home orders) that plausibly created dependence between counties in a state. The independence assumption also implies no interfence: The exposure level $a$ for a given county does not affect the outcomes of the other counties. Given that infectious disease spread does not respect town, county, state, or country boundaries, this assumption may be unrealistic. (2) Consistency: If $A = a$ for any county, then $Y(a) = Y$, and hence the full observed set of outcomes when $A^d = A$ is simply $Y(A^d) = Y$. This means that the counterfactual outcome for a county with its observed mobility level is the observed outcome. This assumption is implied by our definition of counterfactuals as derived quantities through interventions on the NPSEM. (3) Time-ordering: The confounders $W$ precede the mobility exposure $A$, which also precedes the case rate outcome $Y$.

\section{Estimation and inference}
\label{appendix:estimation}

Recall from section \ref{section:Estimation} that we have nuisance function estimates for the conditional outcome $\bar{Q}$ and the density ratio $\Tilde{\lambda}$ using Super Learner; our goal is to estimate $\psi_0(A^d)$. The steps below are modified for the single time point case from the longitudinal version of \cite{diaz_nonparametric_2021}.

\begin{itemize}
    \item Define and calculate weights $\omega_i = \hat{r}(a_i, w_i)$ for $i = 1, ..., n$, calculated from $\Tilde{\lambda}$ estimated via Super Learner
    \item After scaling our bounded continuous outcome $Y$ to $\Tilde{Y} \in [0,1]$ \cite{gruber_targeted_2010}, fit the logistic regression on the observed (scaled) $\Tilde{Y}$ values:
\begin{equation*}
    \text{logit}(\Tilde{Y}|A, W) = \text{logit}(\bar{Q}(A,W)) + \epsilon 
\end{equation*}
    with weights $\omega_i$, calculating the estimated intercept $\hat{\epsilon}$
    \item Update the $\bar{Q}$ estimates as
\begin{equation*}
\begin{aligned}
    \text{logit}(\Tilde{Y}|A^d, W) &= \text{logit}(\bar{Q}(A^d,W)) + \hat{\epsilon}\\
    \hat{\Tilde{Y}}^d &= \text{logit}^{-1}\left[\text{logit}(\bar{Q}(A^d,W)) + \hat{\epsilon}\right]
\end{aligned}
\end{equation*}
\end{itemize}

After unscaling, with the updated estimates $\hat{Y}^d$ in hand, define $\hat{\psi}_{tmle}$ as:
\begin{equation*}
    \hat{\psi}_{tmle} = \frac{1}{n}\sum_{i=1}^n \hat{Y}^d_i.
\end{equation*}

The TMLE estimator is `doubly robust' in that it is consistent if either the estimators of $\bar{Q}$ or $\Tilde{\lambda}$ are consistently estimated and converge at appropriate rates. Further, if both meet those conditions, along with certain regularity conditions, the TMLE converges to a normal distribution with mean 0 and variance equal to the non-parametric efficiency bound, as shown in Theorem 3 of \cite{diaz_nonparametric_2021}. While convergence at the required $n^{-1/4}$ rate for each nuisance parameter is not guaranteed, the inclusion of a wide variety of algorithms in our Super Learner ensemble maximizes our chance of achieving these rates \cite{laan_super_2007}. The efficiency bound for this TMLE depends on the variance of the outcome conditional on $A$ and $W$, the amount of treatment effect variability, and the intensity of shift applied, as measured by the density ratio $r(a,w)$ \cite{diaz_nonparametric_2021}.

Using the empirical mean to estimate $\mathbb{E}_{P_0}(Y)$, we hence have our desired estimate

\begin{equation*}
    \hat{\psi}_{\Delta} = \hat{\psi}_{tmle} - \frac{1}{n}\sum_{i=1}^n Y_i.
\end{equation*}

From this result, variance estimation for $\hat{\psi}$ can be calculated using the estimated influence curves from $\hat{\psi}_{tmle}$ and the empirical mean:

\begin{equation*}
    Var(\hat{\psi}_{\Delta}) = \frac{\frac{1}{n}\sum_{i = 1}^n (IC(O_i) - (Y_i - \bar{Y}))^2}{n},
\end{equation*}

with the influence curve defined as follows: $IC(O) = r(A, W)[Y-\mathbb{E}\left[ Y \mid A, W \right]] + \mathbb{E}\left[ Y \mid A^d, W \right] - \mathbb{E}\left[ Y(A^d) \right]
$\cite{diaz_nonparametric_2021}. This allowed for the calculation of the 95\% Wald-style confidence interval $\hat{\psi}_{\Delta} \pm 1.96 \sqrt{Var(\hat{\psi}_{\Delta})}$.

\clearpage
\newpage

\section{Sensitivity analyses}
\label{appendix:sensitivity}

\subsection{Using one-week leading case rates as outcome}
\label{appendix:sensitivityoneweekY}

\begin{figure}[h]
\begin{center}
\includegraphics[width=\textwidth]{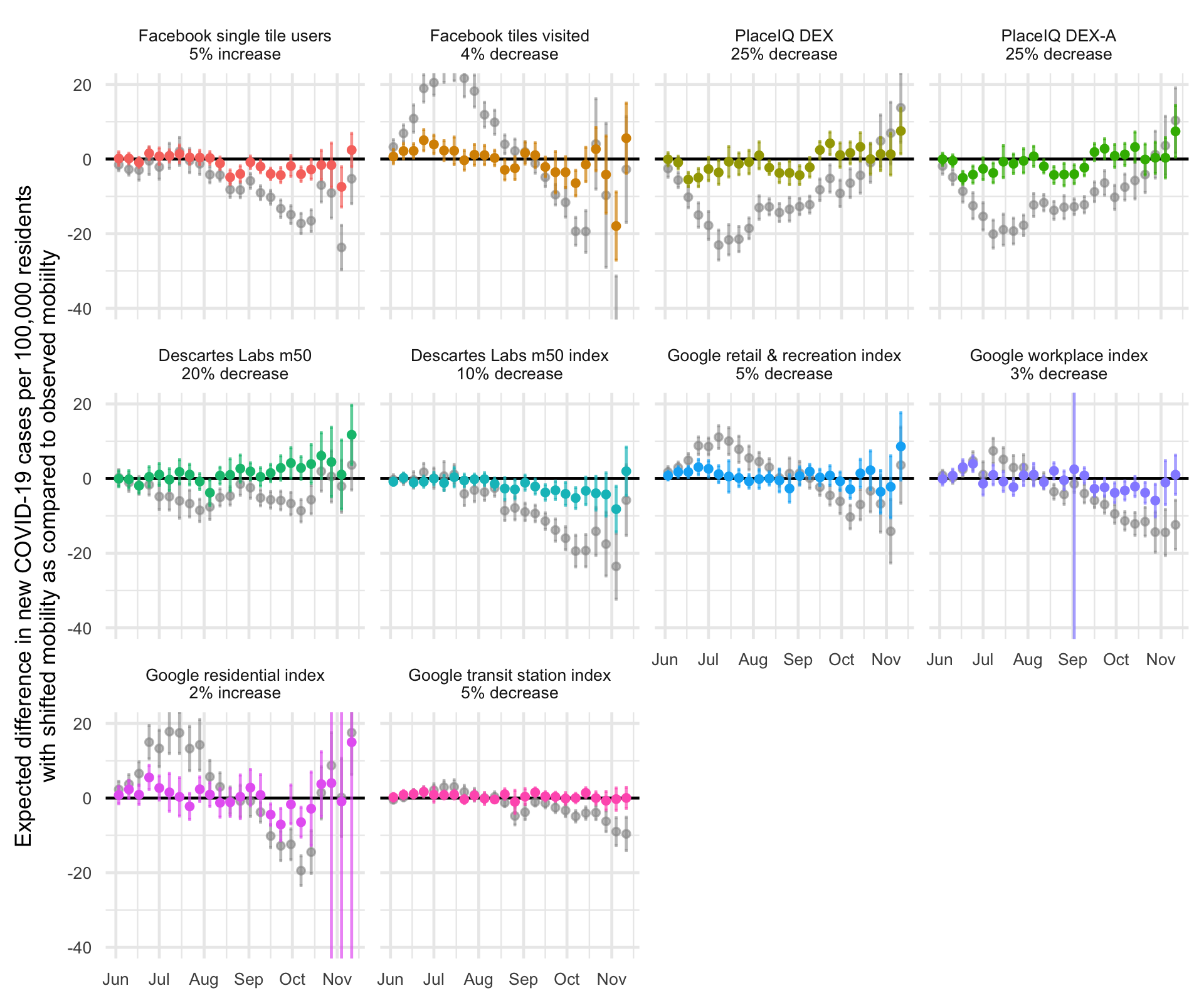}
  \label{fig:1wkslag1}
\end{center}
\end{figure}

\clearpage
\newpage

\subsection{Using current-week cases in the adjustment set instead of one-week lagged cases}
\label{appendix:sensitivitycurrent}

\begin{figure}[h]
\begin{center}
\includegraphics[width=\textwidth]{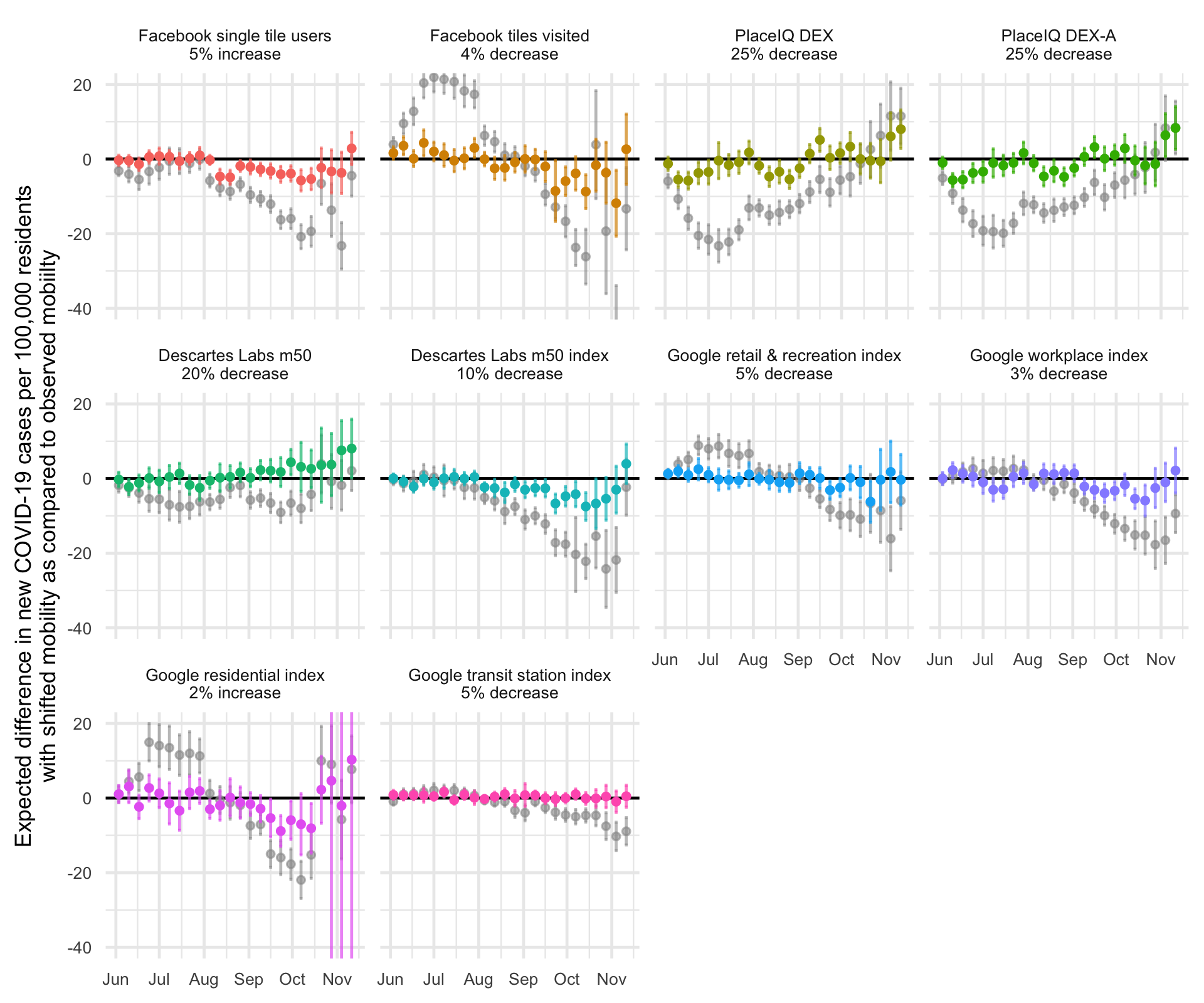}
  \label{fig:2wkscurrent}
\end{center}
\end{figure}

\clearpage
\newpage

\subsection{Observed exposure distributions over time}
\label{appendix:exposureviolins}

\begin{figure}[h]
\begin{center}
\includegraphics[width=\textwidth]{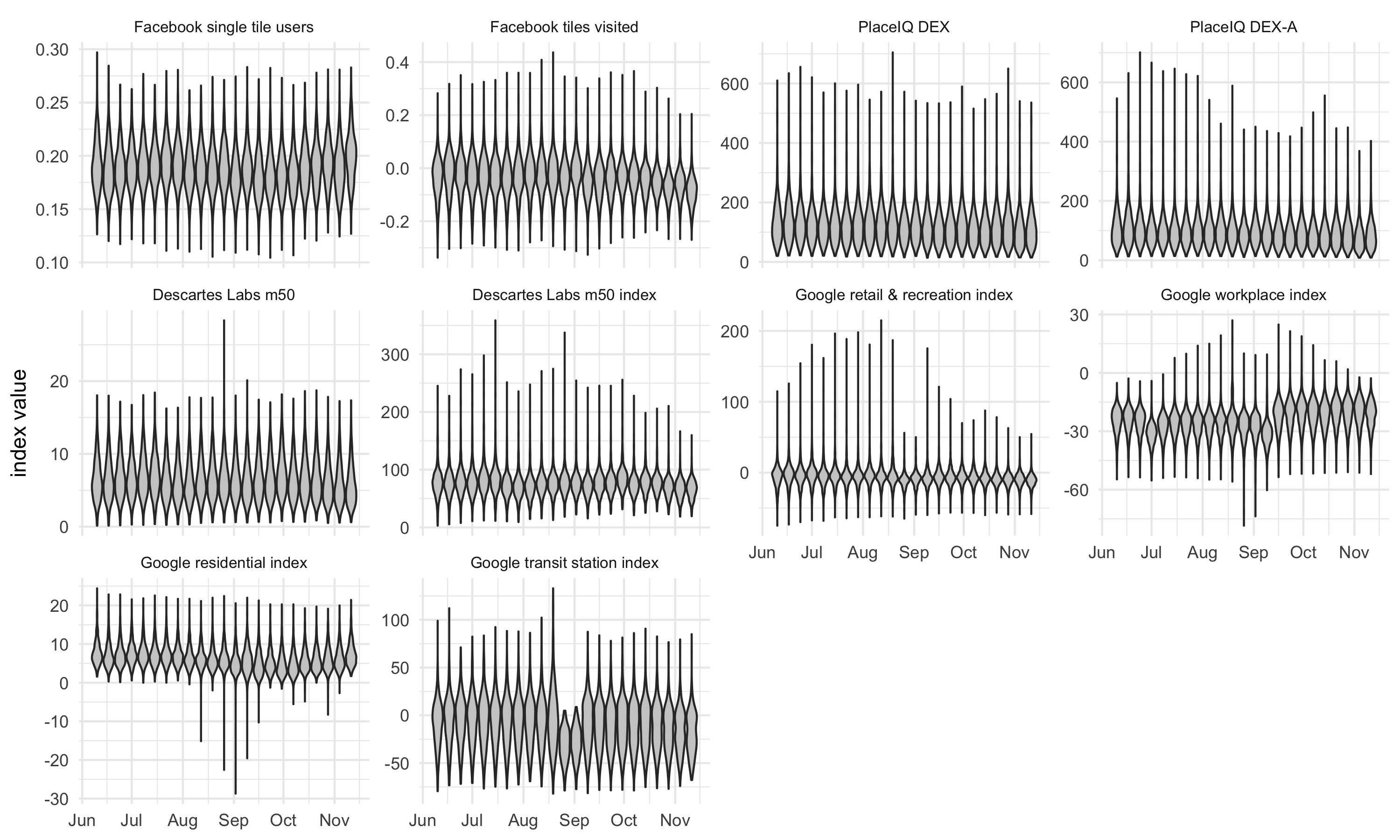}
  \label{fig:exposureviolins}
\end{center}
\end{figure}

\clearpage
\newpage

\section{Supplementary analyses}
\label{appendix:supp}

\subsection{Using a time-varying confounder set}
\label{appendix:supptv}

We tested three of the indices allowing for a different confounder set $W$, screened in each week time-slice by univariate correlations with the exposure and outcome. As with the main analysis, the adjusted results (shown below for several shift value possibilities) show inconsistent patterns over time and CIs frequently overlap with zero.

\begin{figure}[h]
\begin{center}
\includegraphics[width=.85\textwidth]{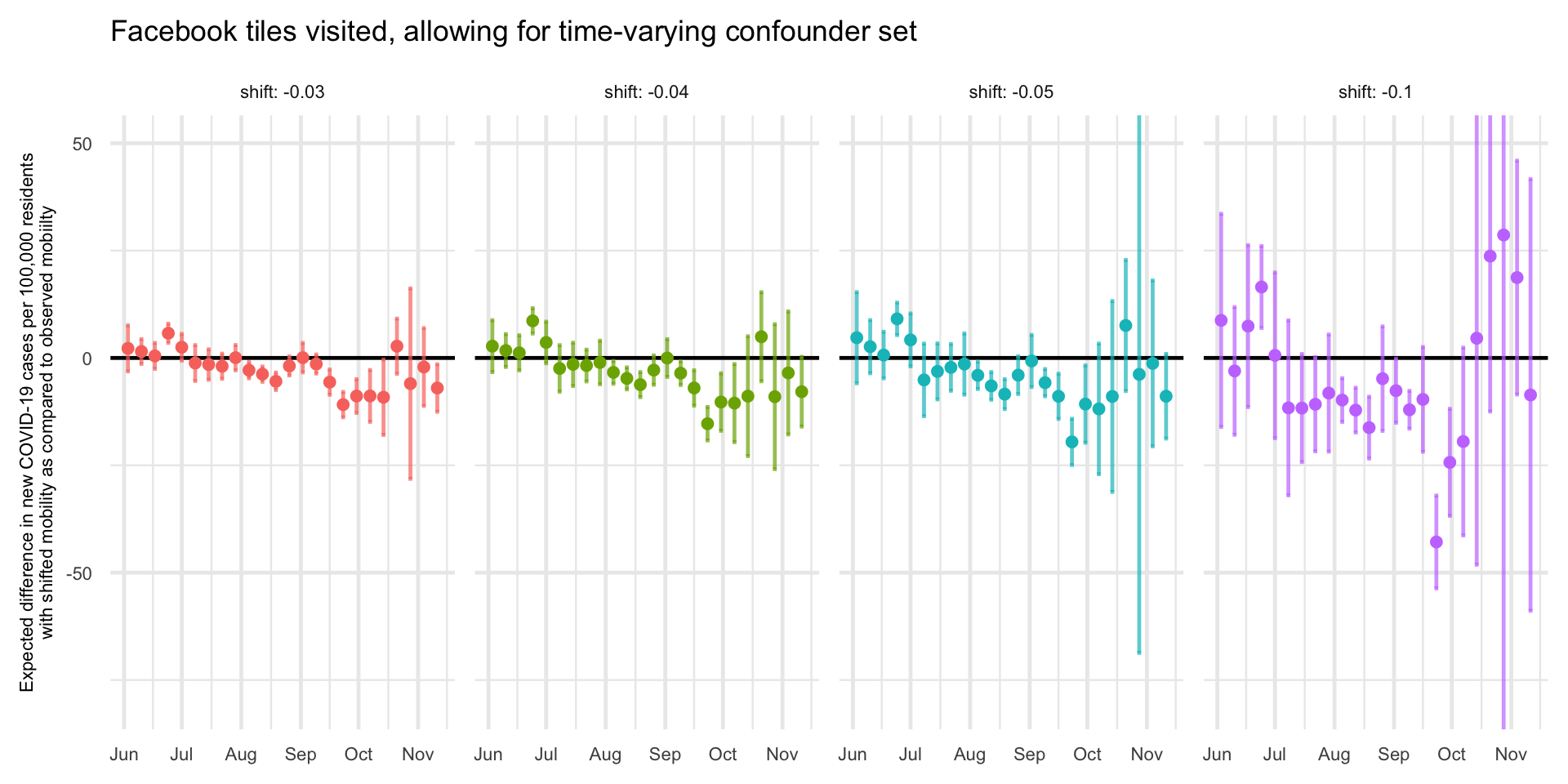}
\end{center}
\end{figure}

\begin{figure}[h]
\begin{center}
\includegraphics[width=.85\textwidth]{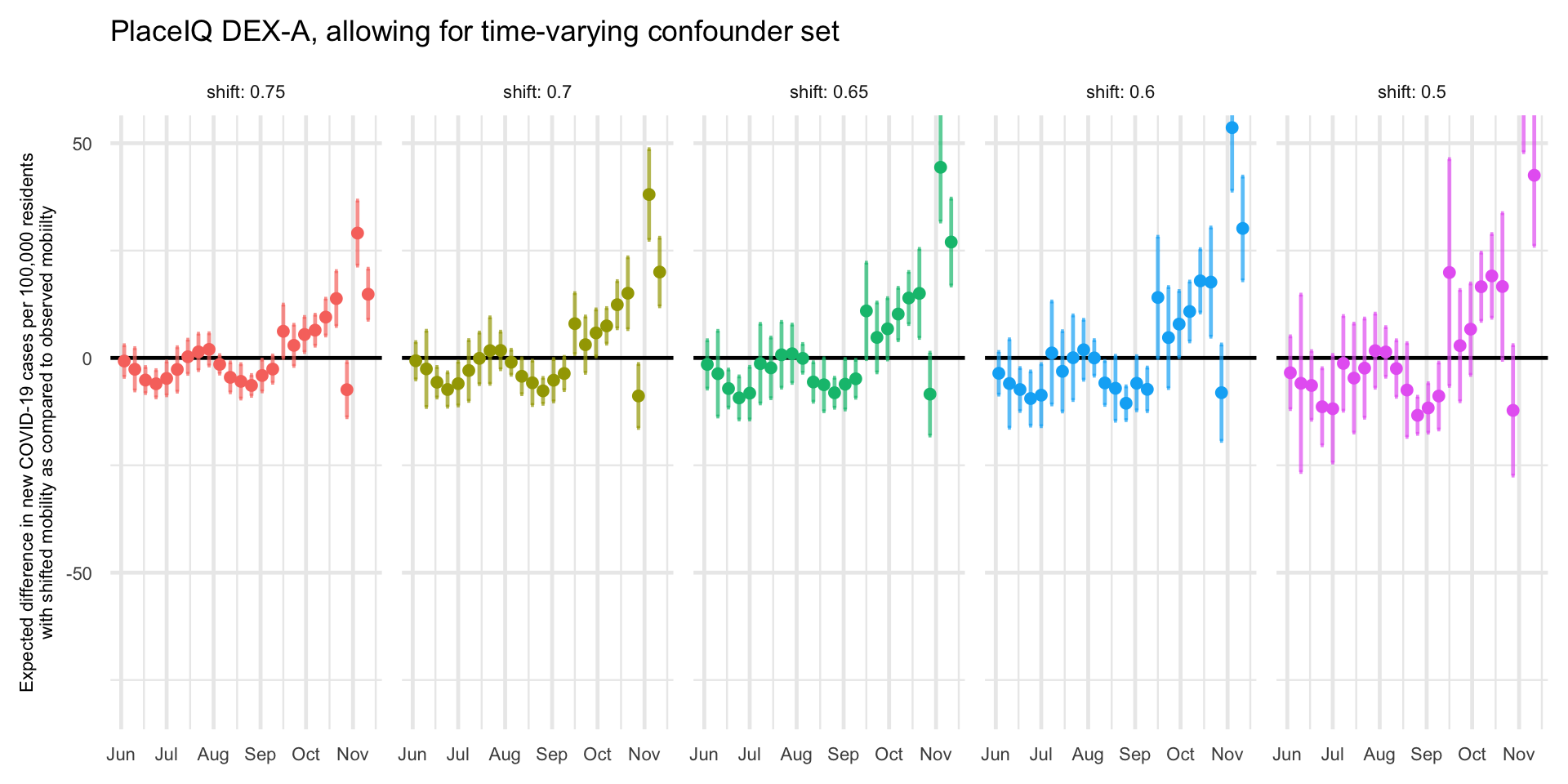}
\end{center}
\end{figure}

\begin{figure}[h]
\begin{center}
\includegraphics[width=.85\textwidth]{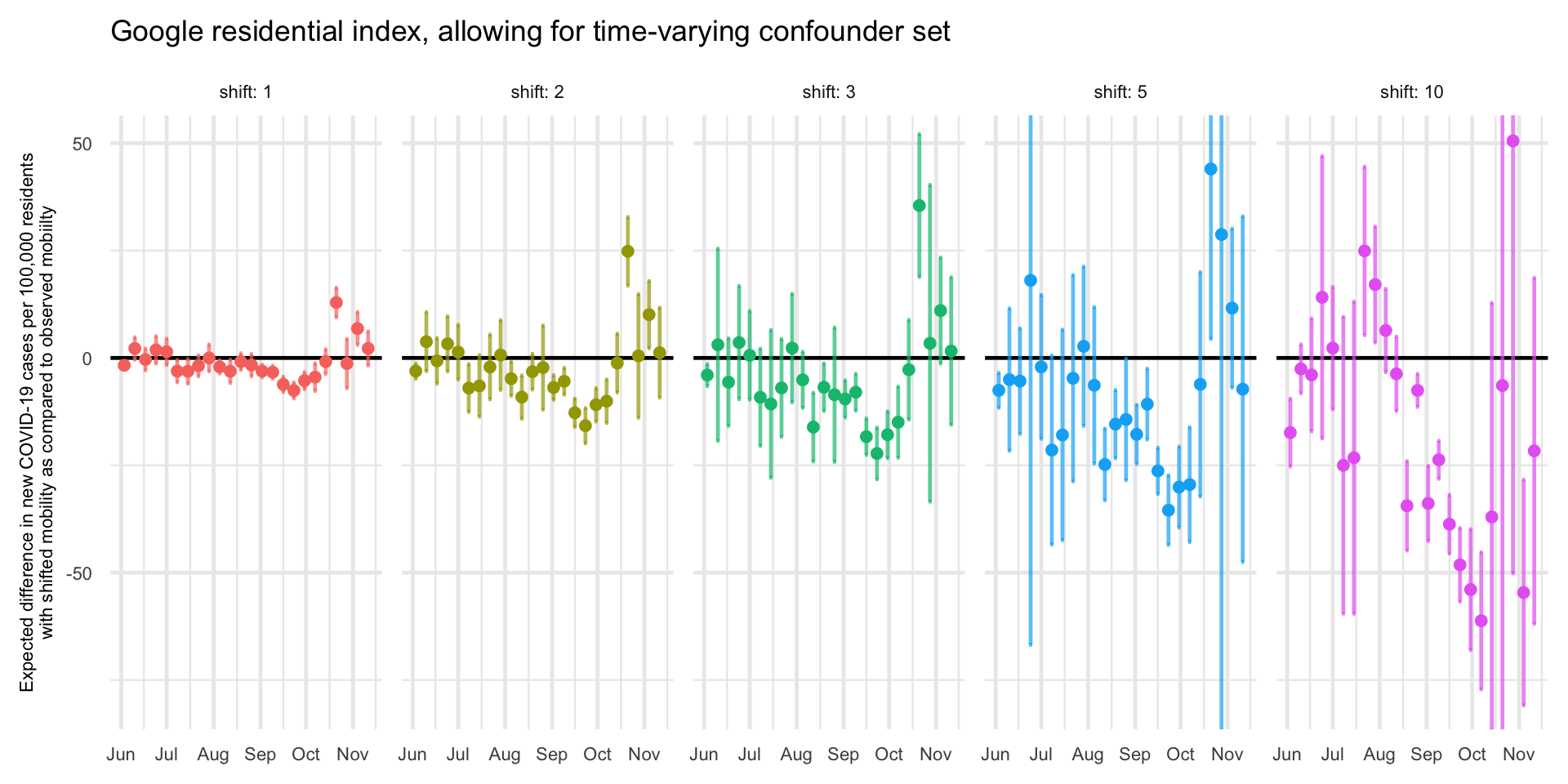}
\end{center}
\end{figure}

\clearpage
\newpage

\subsection{Using a smaller confounder set with larger shifts}
\label{appendix:suppas}

We tested three of the indices with a smaller, time-varying confounder set $W$ (four rather than eight), screened in each week time-slice by univariate correlations with the exposure and outcome, with larger shift values. As with the main analysis, the adjusted results (shown below for several shift value possibilities) show inconsistent patterns over time and CIs frequently overlap with zero.

\begin{figure}[h]
\begin{center}
\includegraphics[width=.85\textwidth]{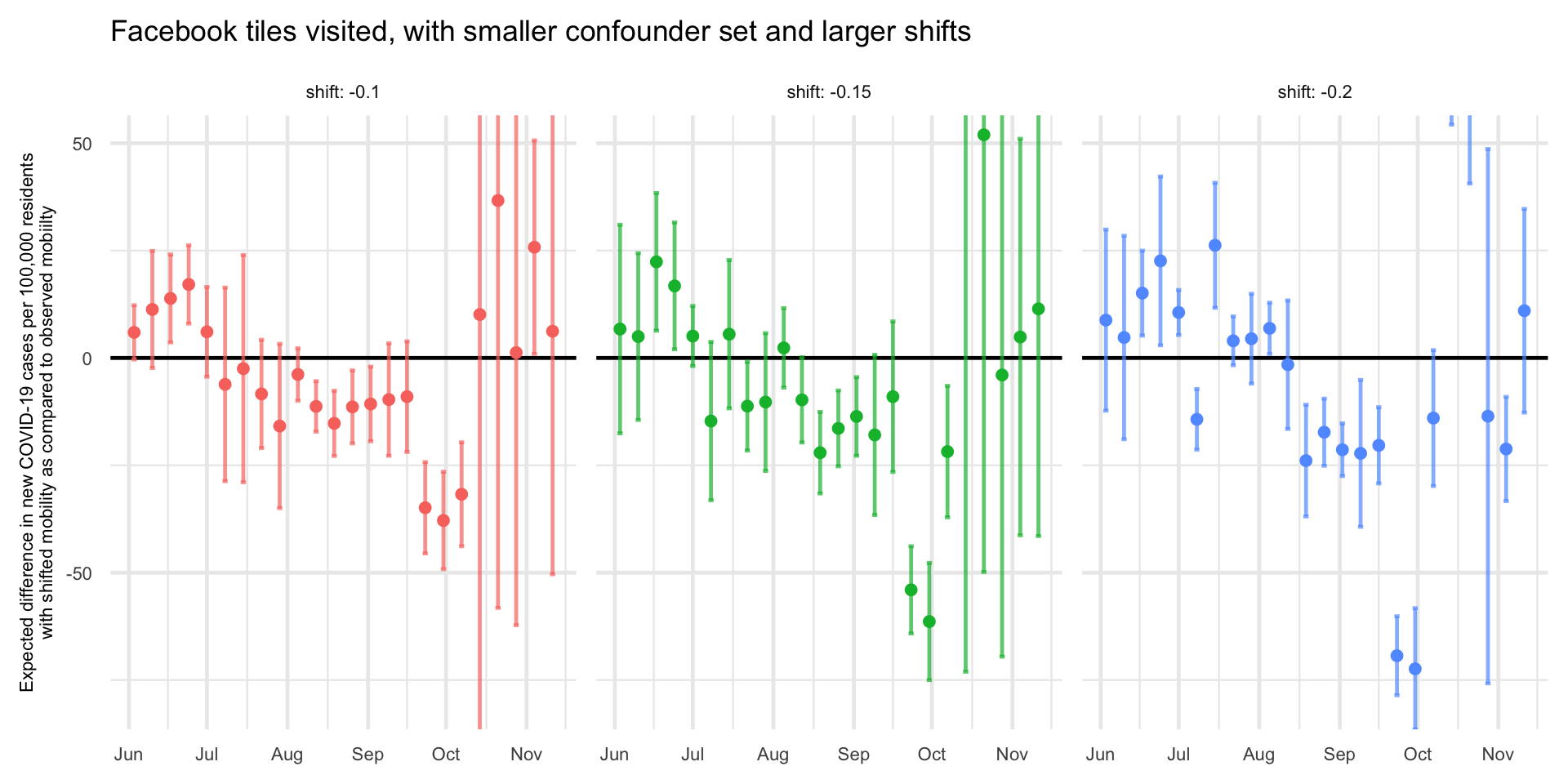}
\end{center}
\end{figure}

\begin{figure}[h]
\begin{center}
\includegraphics[width=.85\textwidth]{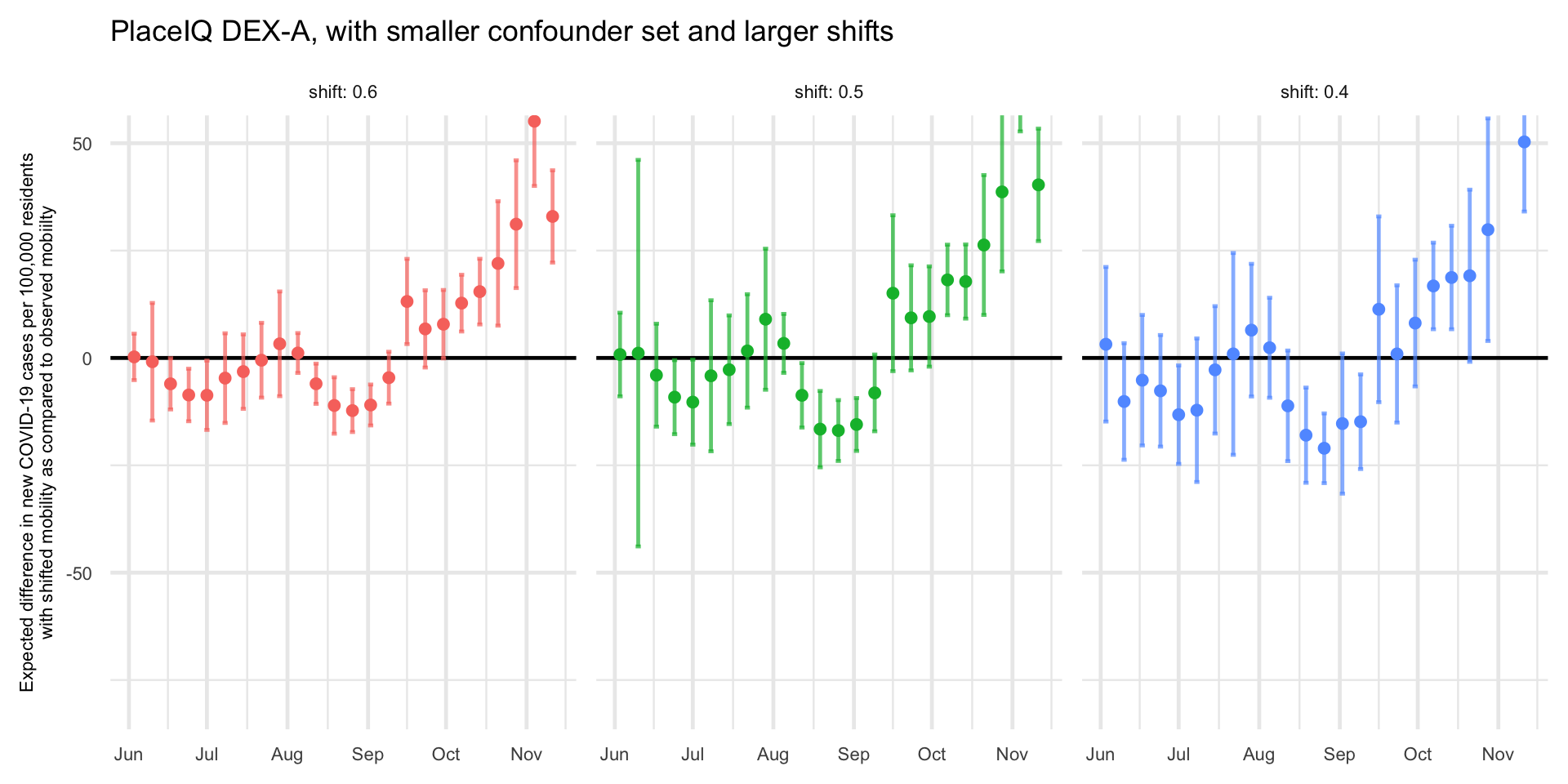}
\end{center}
\end{figure}

\begin{figure}[h]
\begin{center}
\includegraphics[width=.85\textwidth]{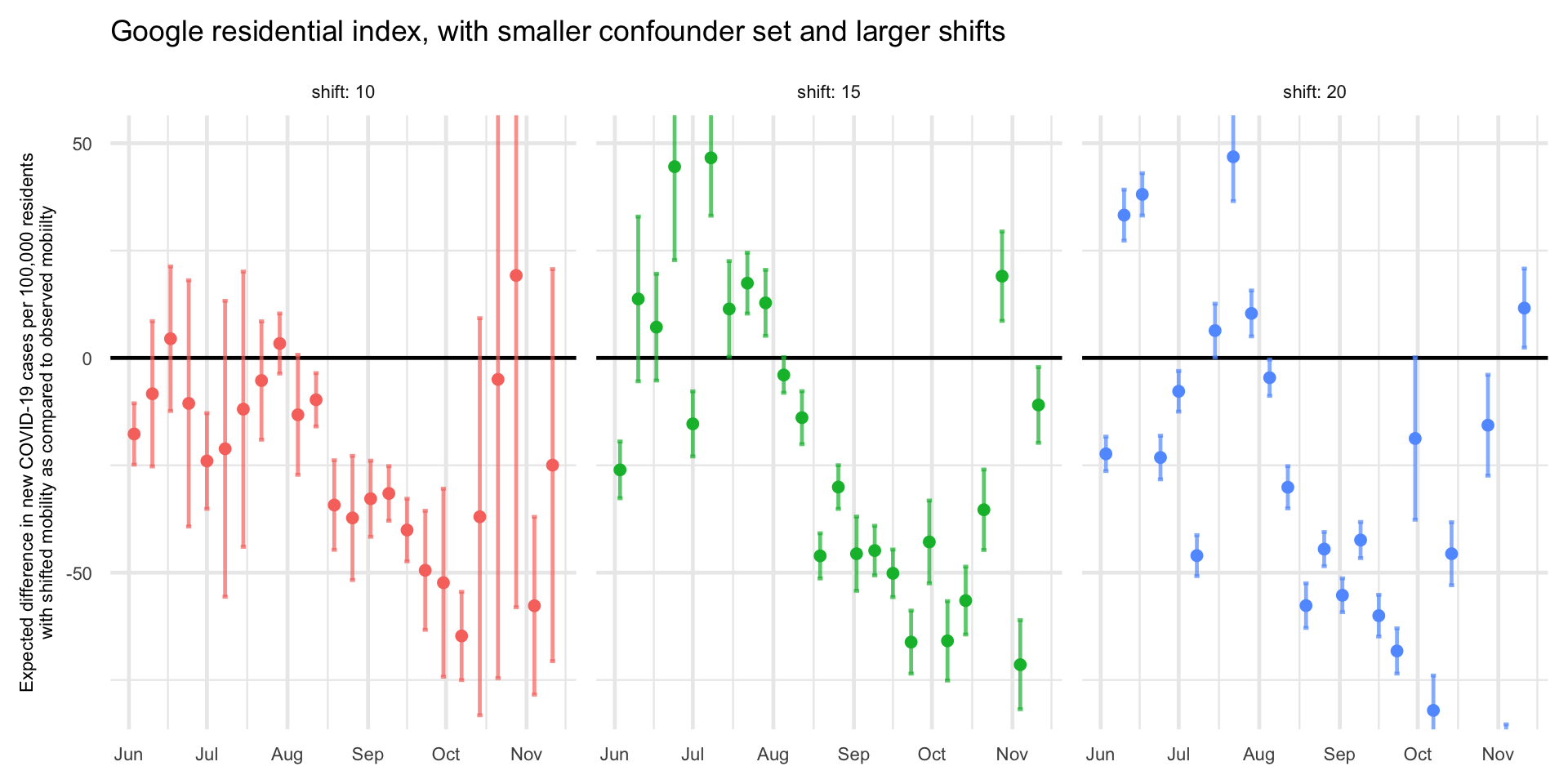}
\end{center}
\end{figure}

\end{appendices}

\clearpage


\end{document}